\documentclass[angeo]{copernicus}

\begin{document}

\title{Diagnostic of the temperature and differential emission measure (DEM) based on \emph{Hinode}/XRT data}

\author[1]{M. Siarkowski}
\author[2]{R. Falewicz}
\author[1]{A. K\c{e}pa}
\author[2]{P. Rudawy}

\affil[1]{Space Research Centre, Polish Academy of Sciences, 51-622
Wroc{\l}aw, ul. Kopernika 11, Poland}

\affil[2]{Astronomical Institute, University of Wroc{\l}aw, 51-622
Wroc{\l}aw, ul. Kopernika 11, Poland}

%% The [] brackets identify the author to the corresponding affiliation, 1, 2, 3, etc. should be inserted.

\runningtitle{Diagnostic of the temperature and differential
emission measure base on....}

\runningauthor{M. Siarkowski et al.}

\correspondence{M. Siarkowski\\ (ms@cbk.pan.wroc.pl)}

\received{}
\pubdiscuss{} %% only important for two-stage journals
\revised{}
\accepted{}
\published{}

%% These dates will be inserted by the Publication Production Office during the typesetting process.

\firstpage{1}

\maketitle

\begin{abstract}
We discuss here various methodologies and an optimal strategy of the
temperature and emission measure diagnostics based on \emph{Hinode}
X-Ray Telescope data. As an example of our results we present
determination of the temperature distribution of the X-rays emitting
plasma using filters ratio method and three various methods of the
calculation of the differential emission measure (DEM). We have
found that all these methods give results similar to the two filters
ratio method. Additionally, all methods of the DEM calculation gave
similar solutions. We can state that the majority of the pairs of
the \emph{Hinode} filters allows one to derive the temperature and
emission measure in the isothermal plasma approximation using
standard diagnostic based on two filters ratio method. In cases of
strong flares one can also expect well conformity of the results
obtained using a Withbroe - Sylwester, genetic algorithm and
least-square methods of the DEM evaluation.
\end{abstract}

%\Keywords {Solar Physics, Astrophysics and Astronomy (Corona and
%transition region; Ultraviolet emission; X rays and gamma rays;
%Instruments and techniques)}

\introduction
%% \introduction[modified heading if necessary]

The solar corona with temperatures greater than a million degrees
emits mainly in X-ray domain of the spectrum. Its emission contains
both the continuum from thermal plasma and line emission of highly
ionized elements. The earliest phases of solar flares are dominated
by non-thermal emissions (hard X-rays), the remainder of the events
manifests itself primarily by thermal emissions at wavelengths
ranging from X-rays through visual. The thermal radiation comes from
plasma having a wide range of temperatures.

Temperature and emission measure diagnostic of the active region and
flaring loops is very important to model the loop structures, and
analysis the heating sources and their (heating) distribution along
the loop.

For the optically thin emission, assuming isothermal plasma
temperature, the well known method of filter ratio, based on ratio
of the observed fluxes in two selected energy (or wavelengths) bands
(\citet{Vaiana}, \citet{Thomas}) can be used. In the case of
multi-temperature plasma  the filter ratio method  derives the
line-of-sight average isothermal temperature and emission measure.
Therefore, in this approach it is more adequate to use many filters
or line in terms of differential emission measure distributions
$DEM(T)/dT$ (\citet{Pottash}, \citet{Sylwester}). Extended
discission of the DEM versus isothermal approximation can by found
e.g. in \citet{Schmelz}, \citet{Cirtain} and references therein.

\begin{figure}[t]
\vspace*{2mm}
\begin{center}
  \includegraphics[width=7.5cm]{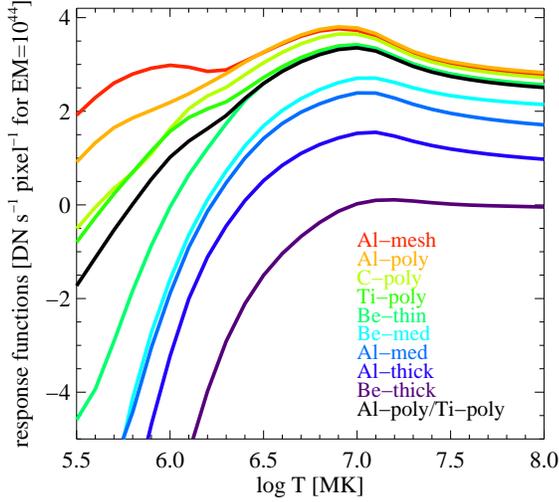}
\end{center}
\vspace{-0.7cm}
  \caption{Theoretical temperature response functions of the selected XRT filters.}
  \label{fig01}
%\vspace{-0.4 cm}
\end{figure}

{\it Hinode} is a new advanced optical, UV and X-ray solar satellite
launched on 22 September 2006. It is equipped with three
instruments: Solar Optical Telescope (SOT), EUV Imaging Spectrometer
(EIS) and the X-Ray Telescope (XRT).  XRT is a high resolution
Wolter-I type grazing incidence telescope, which is sensible to the
emission of plasma in the temperature range 5.5 $<$ log \emph{T} $<$
8. Such wide temperature coverage is realized by nine X-ray filters,
each with its own passband and therefore a different response to
plasma temperature. These filters are mounted on two filter wheels
which allow many combinations of filters and thus can be used for a
temperature diagnostic. Figure~\ref{fig01}  presents the theoretical
temperature response functions of the ten XRT filters combination
for unit volume emission measure of $10^{44}\, \rm{cm}^{-3}$ for
each filter used in this paper. XRT has 2048x2048 back-illuminated
CCD array with $13.5\, \mu$m pixel size. It allows the Sun soft
X-ray images to make in 34 x 34 arcmin FOV with spatial resolution
of ~1 arcsec. Exposure times are in range from 1 ms to 16 sec.
Detailed description of the instrument can be found in the paper by
\citet{Golub07}.

In this paper we will check if XRT response characteristics allow
diagnostics of temperature (\emph{T}) and emission measure
(\emph{EM}) of the coronal plasma. We also describe the observations
selected for the analysis. Additionally we present and discuss
filters ratio method in isothermal approximation and three different
methods for differential emission measure (DEM) analysis in
multitemperature approximation, respectively.

In all calculations we used the temperature response function as
given by SolarSoft and presented in Figure~\ref{fig01}.

\section{Data analysis}

For our work we selected three typical X-ray emitting active regions
observed by {\it Hinode}/XRT with 6 or more different filters. To
prepare the selected data we used standard SolarSoft routine
\emph{xrt\_prep.pro} which is similar in nature to
\emph{sxt\_prep.pro} (\emph{Yohkoh} software). This routine converts
raw data into index and structure arrays. Additionally, this process
removes several instrumental effects. We could mention some of the
most important, for example: replacement of the near-saturated
pixels for value greater than 2500 DN with 2500 DN, remove
radiation-belt/cosmic-ray and streaks, calibration for read-out
signals, removing of the CCD bias, calibration of the dark current,
normalization of each images for exposure time. Using processed data
we calculated their differential emission measure distributions and
using two filters ratio method, temperature and emission measure
maps in isothermal approximation.

\subsection{Two filters ratio diagnostic}

The filter-ratio method is a technique commonly used for plasma
diagnostics based on data from broad-band X-ray imaging systems.
This method requires analysis of two or more intensity maps (images)
of the investigated structures obtained with various broad-band
filters. It delivers maps of temperatures and line-of-sight emission
integrals or emission measure distributions as a function of
temperature.

For a multi-band instruments like XRT, under assumption that the
emitting plasma is optically thin, the observed flux $F_k$ in filter
$k$ in a single pixel is:
\begin{equation}
\label{r1} F_k=\int_{T} \int_{\lambda} f(T,\lambda) \eta_k(\lambda)
EM(T) d\lambda dT
\end{equation}

where $f(T,\lambda)$ is coronal plasma emissivity as function of the
plasma temperature $T$ and the wavelength $\lambda$, calculated at
distance of 1 AU using CHIANTI spectral code (including both
emission lines and continuum). $\eta_k(\lambda)$ is an effective
area of the instrument which includes the geometric area and the
reflectance of the telescope mirror, the entrance and the focal
plane filter transmissions and the CCD quantum efficiency.
\emph{EM(T)} is a collection of emission measures over a range of
temperatures located along the line of sight. Defining temperature
response function $P_k$(\emph{T}) of filter \emph{k} as a
convolution of the instrument effective area $\eta_k$ and emissivity
of solar plasma $f$:

\begin{equation}
\label{r2} P_k=\int_{\lambda} f(T,\lambda) \eta_k(\lambda) d\lambda
\end{equation}

we can rewrite Equation 1 as:

\begin{equation}
\label{r3} F_k= \int_{T} P_k(T) EM(T) dT \approx \sum_{i}
P_k(T_i)EM(T_i)
\end{equation}

For isothermal approximation this Equation can be written as:
\begin{equation}
\label{r4} F_k=P_k(T)EM(T)
\end{equation}

Similar to \citet{Thomas} using Equation 4 one can calculate ratio
$R$ of the observed fluxes ($F$) in two filters labeled "1" and "2"
for a co-aligned pixel area. This can be written as:
\begin{equation}\label{r5}
    R(T)=\frac{F_1}{F_2}=\frac{P_1(T)}{P_2(T)}
\end{equation}
According to Equation 5, the ratio $R$ of observed fluxes depend on
plasma temperature $T$ only. In the next step, one can calculate the
emission measure $EM(T)=F_k / P_k(T)$.

\begin{figure*}[t]
\vspace*{2mm}
\begin{center}
\includegraphics[width=16cm]{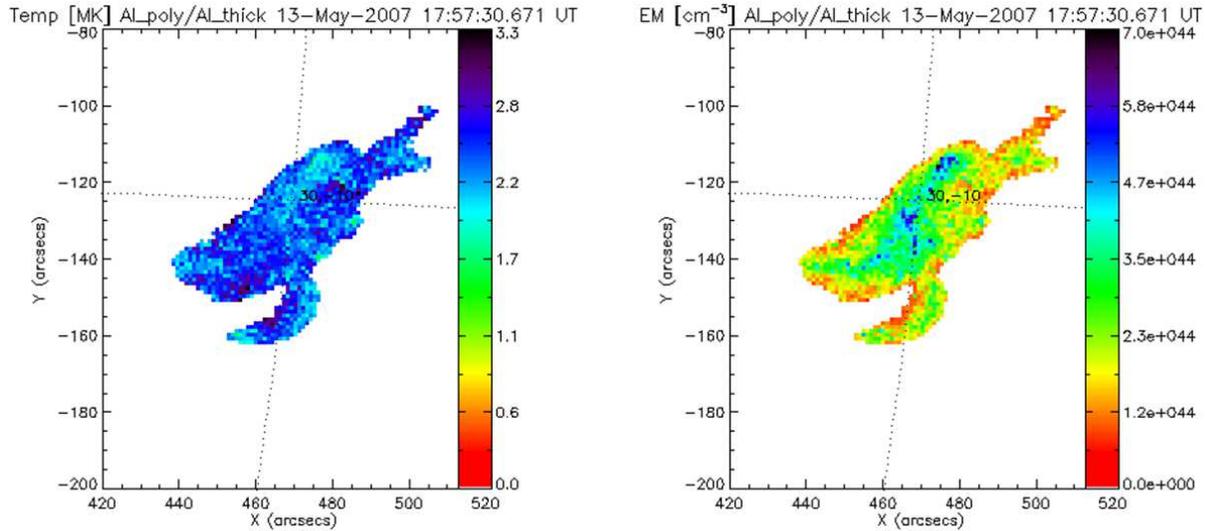}
\end{center}
\vspace{-0.7cm}
\caption{Temperature and emission measure maps
calculated from two filters ratio, Al\_\_poly/Al\_\_thick for the
active region NOAA 0955 observed on 13 May 2007 at 17:57:30~UT.}
\label{fig02}
%\vspace{0.5cm}
\end{figure*}

This procedure can be used on all combinations of the XRT filters,
but it has a limited usability. Especially, ratio $R$ can be
sometimes an ambiguous function in the whole range of temperatures.

Using  procedure described above we made temperature and emission
measure maps for an area limited by an isoline 0.3 of the brightest
pixel on the image.

\subsection{Analysed events}

\begin{figure*}[t]
\vspace*{2mm}
\begin{center}
\includegraphics[width=16.5cm]{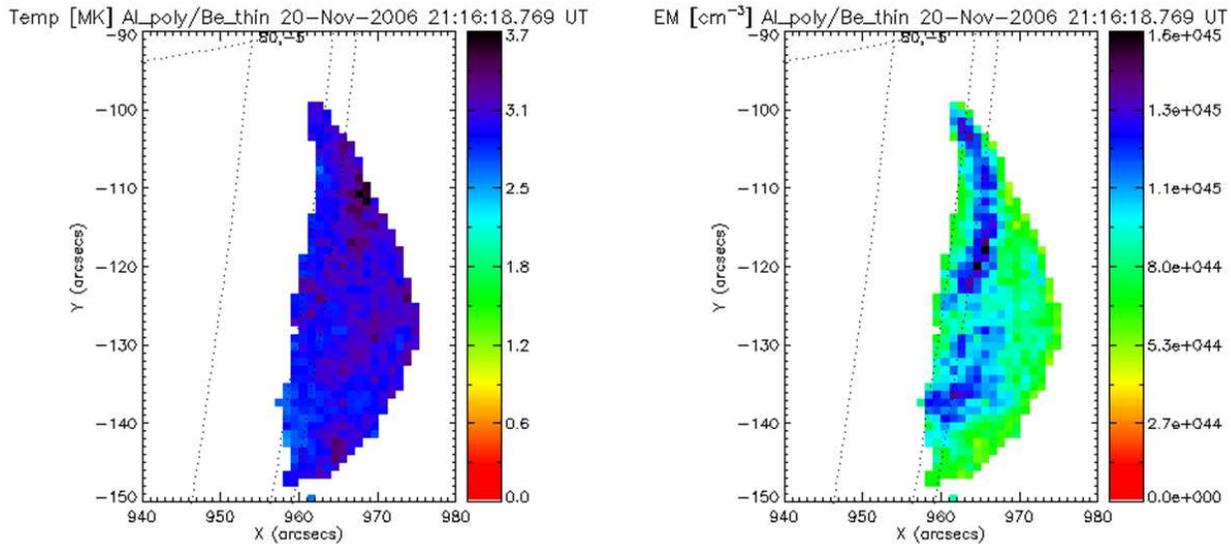}
\end{center}
\vspace{-0.7cm}
\caption{Temperature and emission measure maps
calculated from two filters ratio, Al\_\_poly/Be\_\_thin for the
active region NOAA 0923 observed on West solar limb on 20 November
2006 at 21:16:19~UT.} \label{fig03}
%\vspace{0.5cm}
\end{figure*}

The first analysed region was a relatively faint active region NOAA
0955. Observations were made on 13 May 2007 at around 18:00 UT.
Images taken in ten filters combination are available. Temperature
and emission measure maps were calculated using two pairs of filters
– Al\_\_poly / Al\_\_thick and Ti\_\_poly / Al\_\_thick. As an
example the resulted maps for Al\_\_poly / Al\_\_thick pair are
presented on Figure~\ref{fig02}. For both pairs of filters the
filter ratios are sensitive for temperatures from $log(T)=6.1$ up to
$log(T)=7.7$. However, for all structures of this quiet active
region we obtained temperatures in ranges 2.0-3.3~MK and 1.8-2.4~MK
for the Al\_\_poly / Al\_\_thick and Ti\_\_poly / Al\_\_thick pairs
respectively. The range of the observed temperatures is typical for
solar quiet active regions (see e.g. \citet{Golub97}).

An active region NOAA 0923 was seen on the edge of the West solar
limb between 19 and 23 November 2006. Observations analyzed here
were made on 20 November 2006 at around 21:16 UT. Images taken in
six filters are available. Temperature and emission measure maps
were calculated for two pairs of filters: Al\_\_poly/Al\_\_mesh and
Al\_\_poly/Be\_\_thin. The resulted maps for Al\_\_poly/Be\_\_thin
filters pair are presented on Figure~\ref{fig03}. The highest
temperatures range of 3.5 MK are observed in the brightest loop,
seen just on the solar limb.

\begin{figure*}[t]
\vspace*{2mm}
\begin{center}
\includegraphics[width=16.5cm]{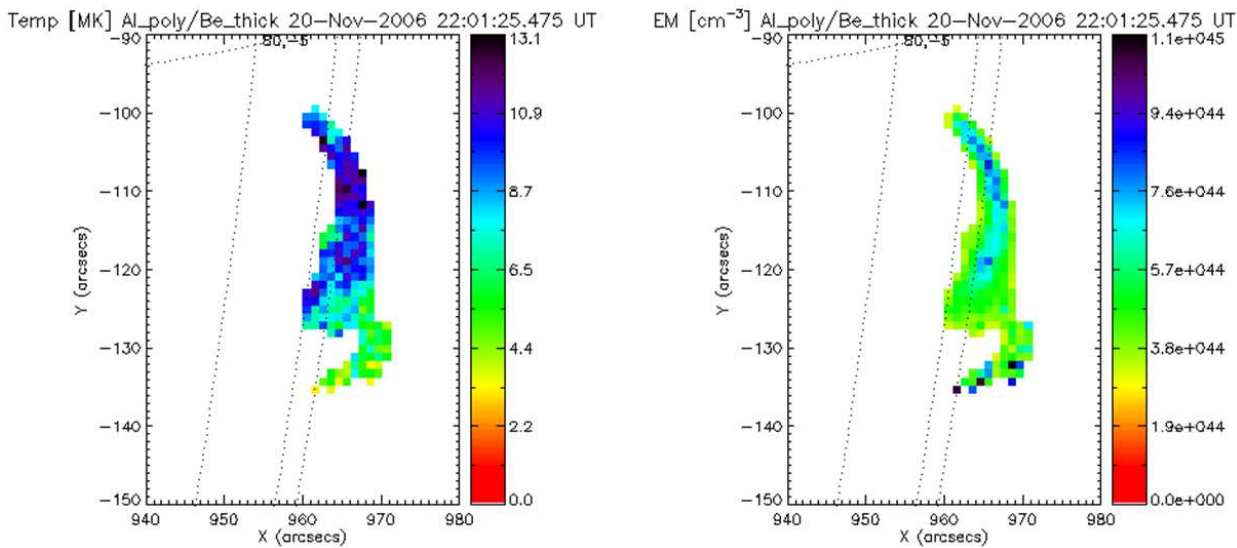}
\end{center}
\vspace{-0.7cm}
\caption{Temperature and emission measure maps
calculated from two filters ratio, Al\_\_poly/Be\_\_thick for the
active region NOAA 0923 observed on West solar limb on 20 November
2006 at 22:01:25~UT.} \label{fig04}
%\vspace{0.5cm}
\end{figure*}

The same active region NOAA 0923 was observed one hour later with
seven filters. Temperature and emission measure maps were calculated
for two pairs of filters: Al\_\_poly/Al\_\_mesh and
Al\_\_poly/Be\_\_thick. The resulted maps for Al\_\_poly/Be\_\_thick
pair are shown in Figure~\ref{fig04}.

Due to the construction of the filter wheels, XRT images in various
filters are only obtained consecutively. Thus an important
assumption underlying the filters ratio diagnostic (and also DEM
analysis described below) is that there was no substantial change in
the morphology and energetics of the event between two images using
various filters. This assumption is usually fulfilled in the case of
quiet Sun and active region observations. Checking the light-curves
for different filters we proved that this assumption is also true in
 the observations above described. Moreover, in the case of small B1.5 flare
images with different filters were obtained first around the maximum
and the second one on decay phase of the event. We use this pair
because the second filter was only one "thick" filter available. The
exact co-alignment of images is likewise important, especially for
the DEM calculations in the individual pixels.

Using 'harder' Al\_poly/Be\_thick filters pair instead of 'softer'
Al\_poly/Be\_thin pair results in obtaining much higher
temperatures. Temperatures in the brightest loop reached average
values up to ~13 MK. Such a high temperature could be connected to
the appearance of a small B1.5 class solar flare observed in this
active region. The flare started at 22:00~UT, reached its maximum at
22:05~UT and ended at 22:09~UT. In fact temperature obtained from
\emph{GOES} data for this small flare reached the value around 10
MK.

It is possible with XRT to register images of reasonable S/N with
cadences 1 or 2 seconds. In the case of solar flares where fluxes
can vary rapidly it is possible in principle to interpolate the
intensities of one image to match the time at which the image in the
other filter was acquired. Examples of temperature and DEM analysis
of solar flare are well know in the case of \emph{Yohkoh}
observations (\citet{McTiernan}). Examples of DEM analysis with
\emph{Hinode}/XRT in the \emph{GOES} C8.2 flare is presented by
\citet{Reeves}.

\subsection{Distribution of the Differential Emission Measure}

Differential emission measure, defined as $DEM(T)=d(EM)/dT=n_e ^2
dV/dT$, is a useful diagnostic tool of non-isothermal plasma. It is
usually calculated from a number of line fluxes formed at different
temperature ranges. Generally the determination of the DEM
distribution is an ill-conditioned problem with no unique solution
\citet{Craig} and this problem needs some kind of regularization.
Calculations of DEM from broand-band filters observations is also
possible if a corresponding filters have different temperature
response functions (see Figure~\ref{fig01}). In the case of {\it
Hinode} broad-band filter observation such method was worked out and
tested by \citet{Weber}.

\begin{figure*}[t]
\vspace*{2mm}
\begin{center}
  \includegraphics[width=7.7cm]{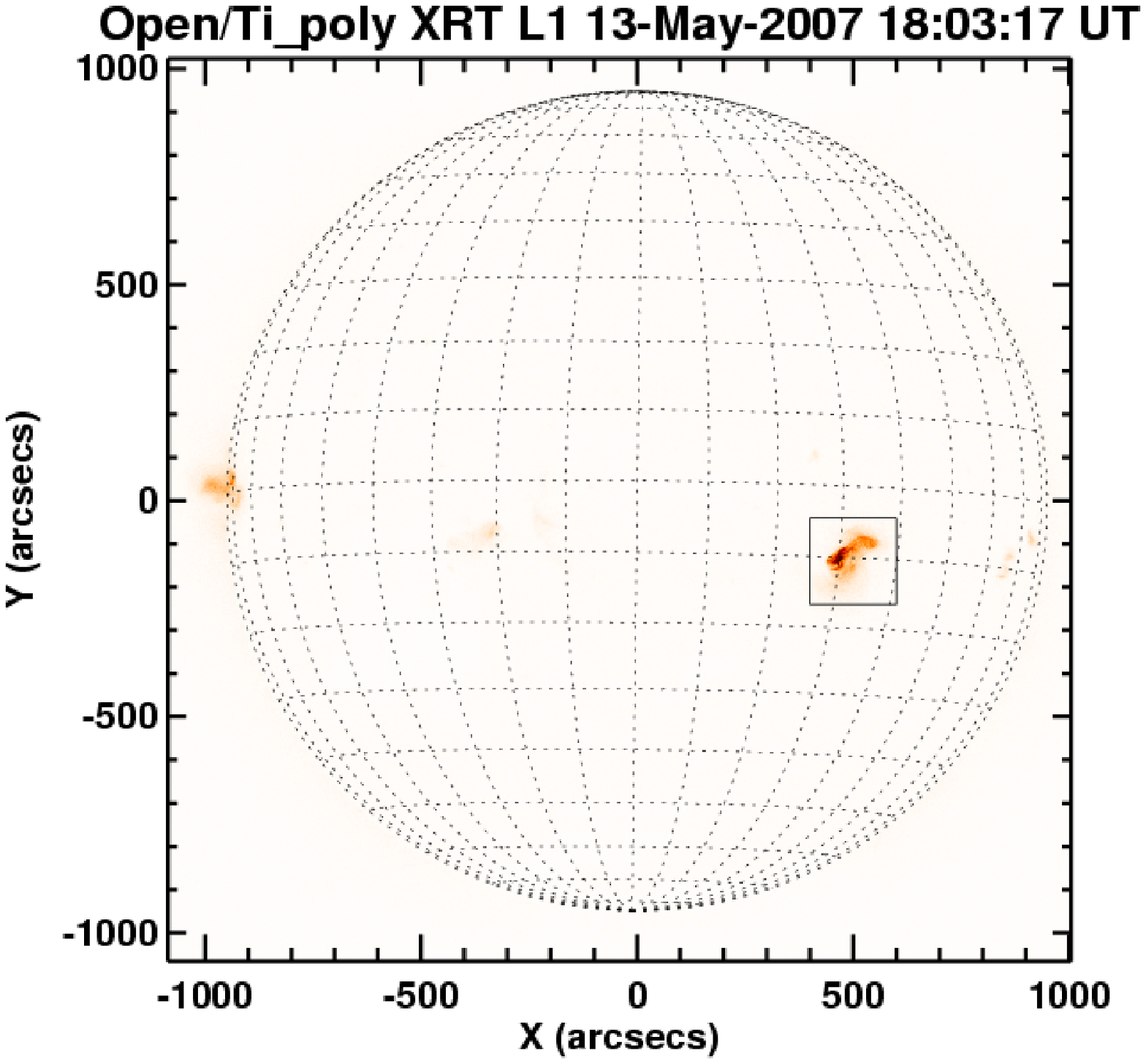}
   \includegraphics[width=7.7cm]{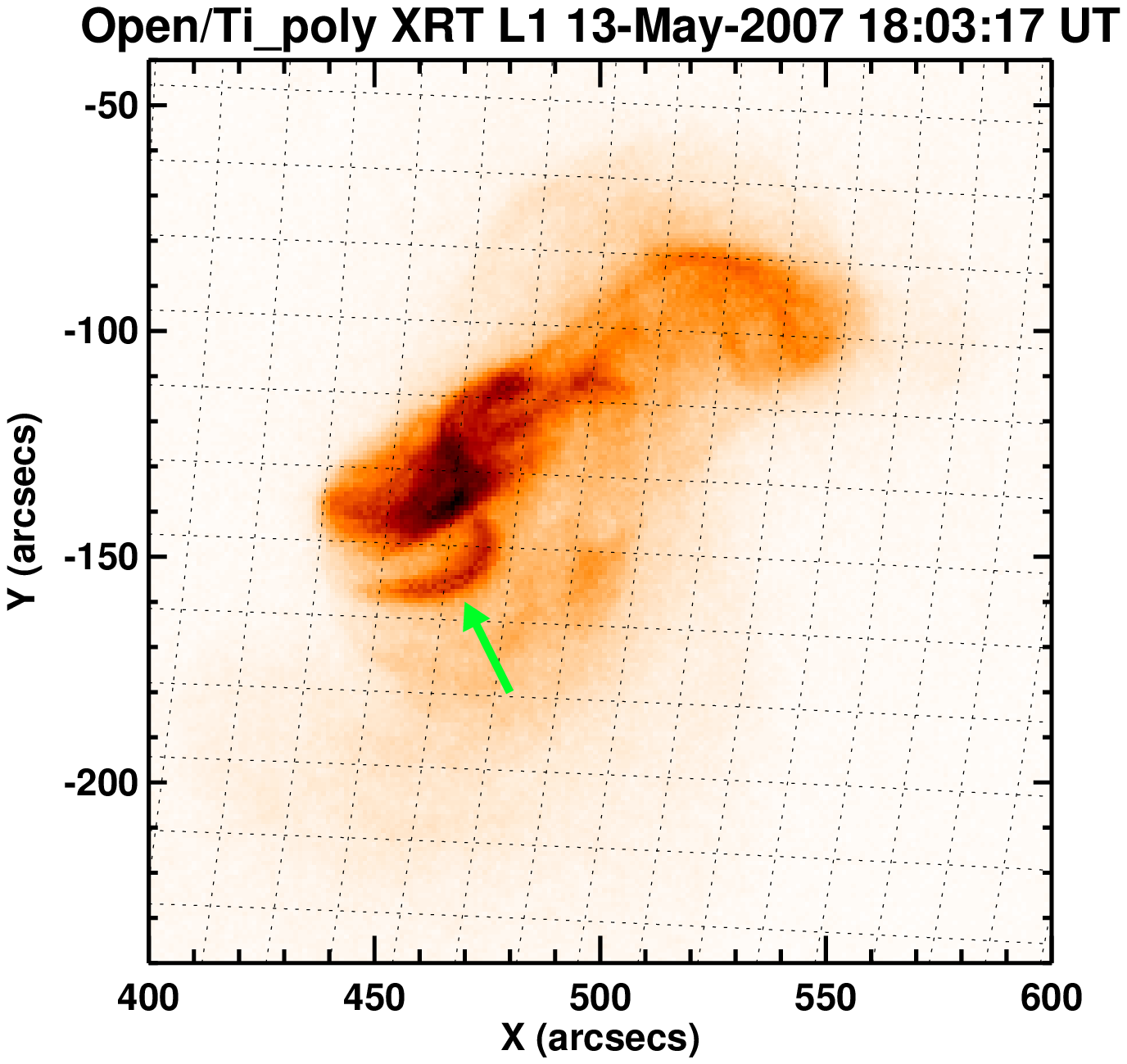}
\end{center}
\vspace{-0.7cm}
\caption{On the left: full Sun image obtained by
XRT/{\it Hinode} with Ti\_\_poly filter. On the right: enlarged map
of NOAA 955 active region observed by {\it Hinode}/XRT telescope
with Ti\_\_poly filter. The DEM distribution, presented on Figure 7,
was calculated for the loop pointed by the green arrow.}
\label{fig05}
%\vspace{0.5cm}
\end{figure*}

In this paper we compare three different methods of the DEM
reconstruction using {\it Hinode}/XRT observations with ten filters.
The methods used are as follows:

Genetic algorithm (GA) - This algorithm is based on ideas of
biological evolution and natural selection mechanism. Starting from
randomly chosen initial populations of different DEMs a new
generation of DEMs is produced by crossover and mutations. Process
of breeding (and multiplication) of the whole population is
controlled by assumed fitness criterion, e.g. DEMs with higher
fitness have a higher probability to participate in the process of
multiplication using crossover. Examples of DEM distribution
calculated with genetic algorithm cam be found for example in papers
by \citet{Kaastra}, \citet{Guedel} and \citet{McIntosh}.

  We used an IDL implementation of Charbonnaue's 'PIKAIA' Fortran code
\citet{Charbonnaue}. Our population had 1000 individual DEM
distributions or 'genes'. After a process of multiplication a new
populations was created consisting of 90\% new DEMs and 10\% of the
old ones with the highest fitness. We used $\chi^2$ as the fitness
criterion of the DEM. The process of evolution was stopped after 10
000 - 20 000 generations, when the convergence became very slow. The
minimum values of $\chi^2$ were around 4.3; typically 15 to 30
'best' newly generated DEMs had $\chi^2$ values below 10. The
evolution process was repeated 6 times, each time starting from a
new random population. As a result we obtained 136 solutions with
$\chi^2 <$ 10. In Figure~\ref{fig07} the green line presents the
mean of these 136 solutions ($\chi^2$ of the mean is 4.7). The
errors were calculated as a standard deviation from the mean. The
solution obtained with GA has $\chi^2$ less than those obtained with
WS and LS methods.

\begin{figure*}[t]
\vspace*{2mm}
\begin{center}
  \includegraphics[width=5.8cm]{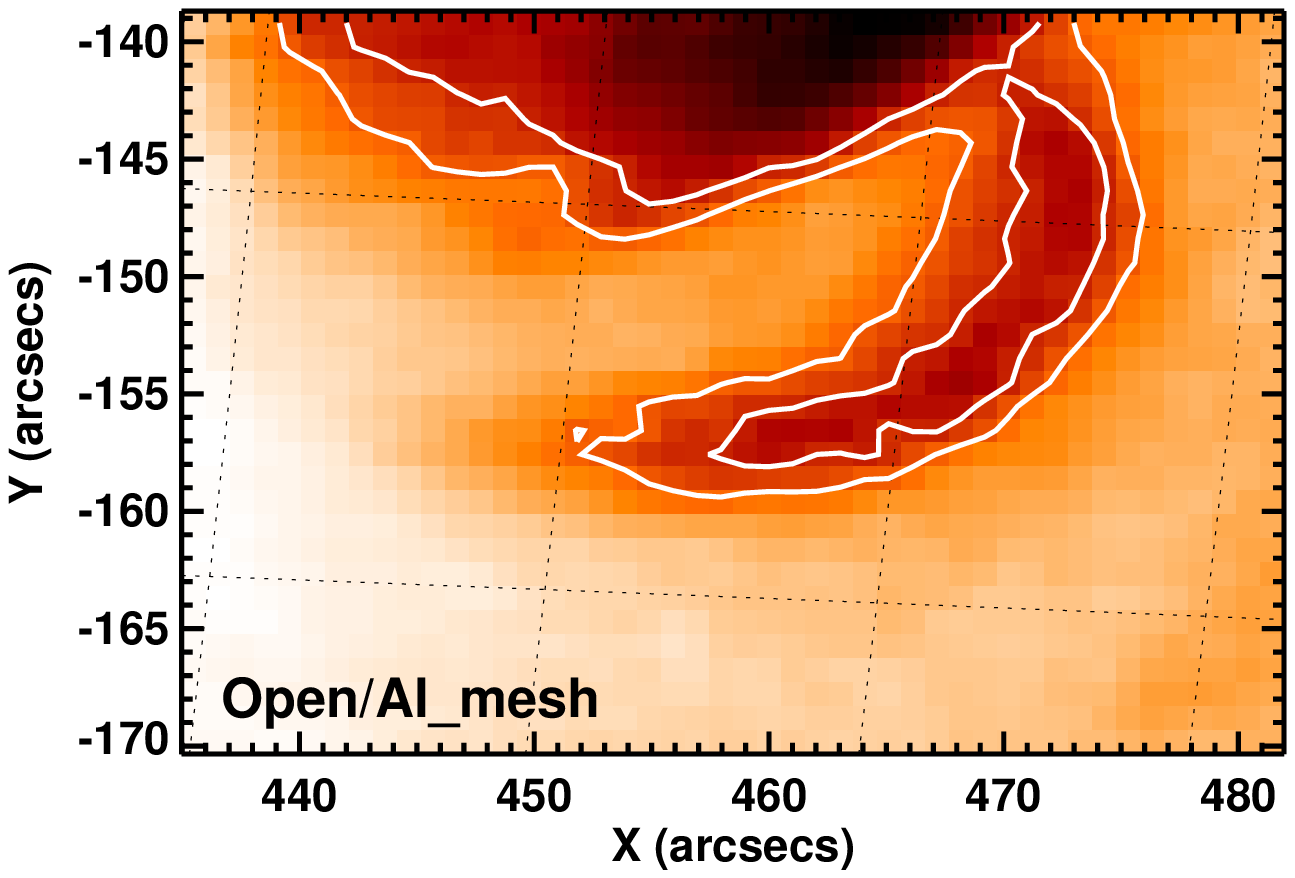}
    \includegraphics[width=5.8cm]{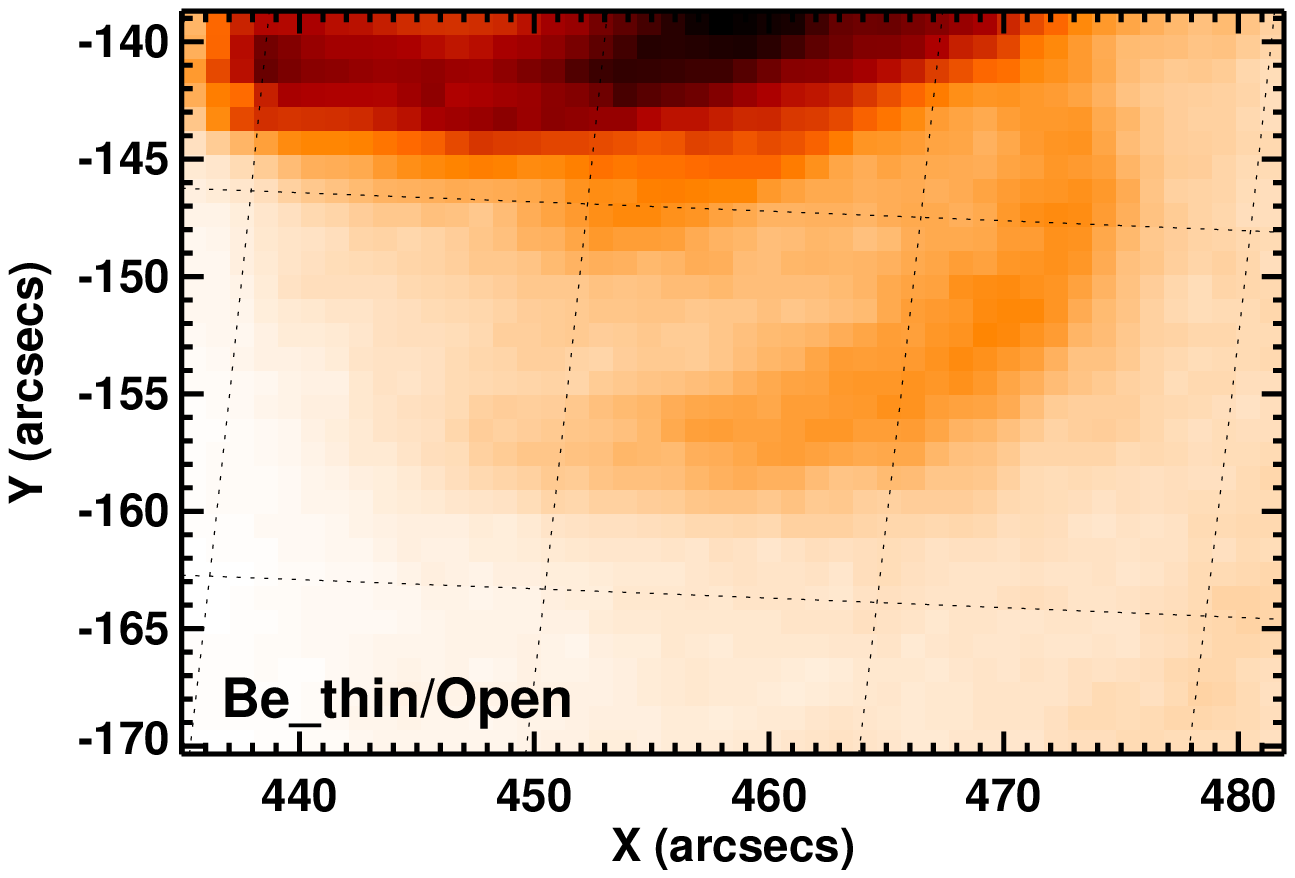}
      \includegraphics[width=5.8cm]{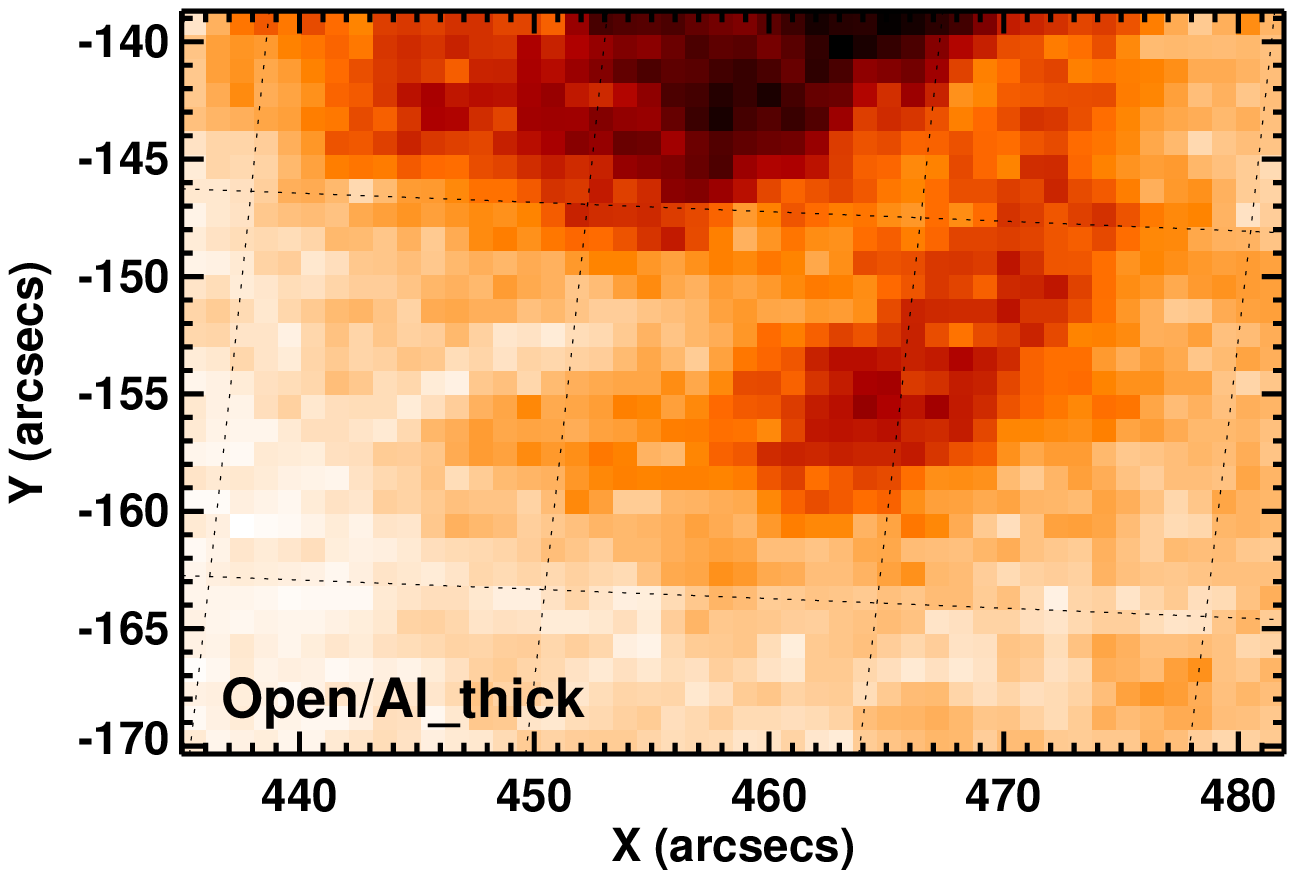}
\end{center}
\vspace{-0.7cm}
  \caption{Images of the loop observed in active region NOAA 09555 around 18:00~UT, taken in three
different {\it Hinode}/XRT filters. Contours overplotted on an
Al\_\_mesh filter image corresponds to 0.6 and 0.8 of the maximum
flux of the loop.}
\label{fig06}
%\vspace{0.5cm}
\end{figure*}

The Withbroe - Sylwester (WS) method is iterative, multiplicative
algorithm relying on the maximum likelihood approach. In our
calculations we started from constant value DEM approximation. For
detailed descriptions of this method see Withbroe (1975) and
Sylwester et al. (1980). Using WS method we performed 200
simulations consisting of 3000 iterations each. The obtained
$\chi^2$ were in the range from 5 to 25. The mean solution has
$\chi^2 = 10$ (see Figure~\ref{fig07}).

Least-square (LS) method - this procedure calculates DEM basing on a
least-squares fit of the calculated fluxes to the observed ones. The
method is described in more detail by \citet{Golub07} and
\citet{Weber} and implemented as a \emph{xrt\_dem\_iterative}
routine in SolarSoft. The DEM function is evaluated from some spline
points, which are directly manipulated by the $\chi^2$ are fitting
the routine. Using LS method we performed 200 LS-simulations. The
obtained $\chi^2$ were in the range from 20 to 30. The mean solution
has $\chi^2 = 25$.

\begin{figure}[t]
\vspace*{2mm}
\begin{center}
\includegraphics[width=8.5cm]{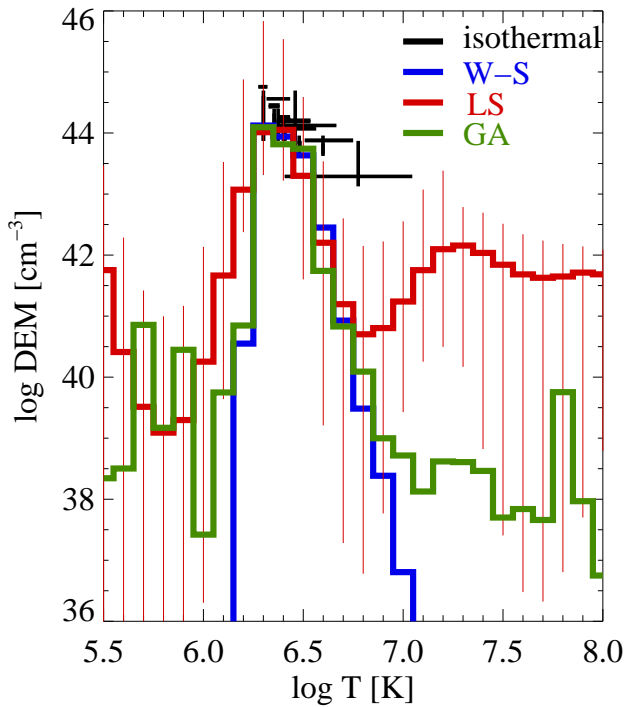}
\end{center}
\vspace{-0.7cm}
\caption{The DEM distributions of the loop observed
on 13th May 2007 obtained using the three analysed numerical methods
and using isothermal approximation with different filter pairs.}
\label{fig07}
%\vspace{0.3cm}
\end{figure}

As an example we have calculated DEM for the loop observed on 13th
May 2007 (see Figures~\ref{fig05} and ~\ref{fig06}). The flux used
in DEM calculation was averaged over an area delimited by 0.8
contour level on the Al\_\_mesh filter image (see
Figure~\ref{fig06}, left panel).

We have tested all three methods using synthetic isothermal model.
The reconstructed width of the solutions allow us to estimate the
temperature resolutions of the method. This resolution is low at the
edges of considered temperatures range (5.5 $<$ log \emph{T} $<$ 8)
and increase in the temperatures where response functions have their
maxima (see Figure~\ref{fig01}) with best values of dlogT = 0.5-0.6,
0.2-0.3 and 0.1-0.2 for LS, WS and GA method respectively. DEM
errors are similar for all three methods, and we showed in
Figure~\ref{fig07} only error bars for LS for readness and
transparency.

The uncertainties are in fact very large at lower temperatures (see
Figure~\ref{fig07}), and for this reason all three methods give
solutions lying within error limits of both other methods. Thus one
can not evaluate precisely \emph{EM} in this temperatures range
using \emph{Hinode}/XRT data. However, it is well known that DEM has
a minimum around log \emph{T} $\approx$ 5.5 for quiet Sun and active
regions and for this reason each obtained solution could be right
(\citet{Brosius}, \citet{Schmelz}).

There are also differences in DEM solutions obtained for high
temperatures. The LS solution indicates the presence of a bigger
amount of the plasma having the temperatures in the range (7.0 $<$
log \emph{T} $<$ 7.3) than the WS and GA methods. The differences
are larger than estimated errors. A possible explanation of these
differences is that LS models depend on initial approximation of DEM
(\citet{Weber}). LS method has also the lowest temperature
resolution comparing to WS and GA methods. The difference between
all three methods should be extensively tested using XRT
observations.

Obtained $\chi^2$ are rather high. Considering whole loop instead of
one or two pixels, the relative errors of flux decrease and thus
$\chi^2$ increase (being much above 1.0). On the other side there
are still large uncertainties in the fluxes measured with
Al\_\_thick filter which strongly influence DEM in high
temperatures. Large $\chi^2$ value can be also attributed to
uncertainness in atomic data which contribute to response functions
and to systematic error in background rejection.

\section{Discussion and conclusions}

XRT telescope can detect coronal plasma of various temperatures
ranging from less than 1\,MK up to more than 10\,MK. Applying
observations made with two or more filters one can analyze coronal
temperature and emission measure along a line-of-sight.

The filter-ratio method has been used frequently in the past, but
the high quality XRT data give temperature and emission measure maps
with unprecedented accuracy and resolution. However, filter-ratio
method is not appropriate for multi-temperature structures, giving
an average value of \emph{T} and \emph{EM} only. For such a case the
DEM methods of analysis of the distribution of emission measure
should be used.

The DEM distributions obtained with the described methods are
presented on Figure~\ref{fig07}. All three methods give the main
peak of emission for log $T = 6.3$. Observed differences between
results of the three methods lie more or less within error limits
(please note the logarithmic scale of DEM). This figure presents
also a set of isothermal solutions obtained for different pairs of
filters. Their values lie around the main peak of DEM obtained with
numerical methods described before. Thus, we can conclude that all
three methods give results similar to the two filters ratio method.

We can also conclude that described methods of the DEM calculation
gave similar and comparable solutions. Additionally, based on shapes
of the emission functions (see Figure~\ref{fig01}) we can expect
much better conformity of the results obtained with WS, LS and GA
methods in a case of strong flares. We can state that the majority
of the pairs of the {\it Hinode} filters allows one to derive the
temperature and emission measure in the isothermal plasma
approximation using standard diagnostic based on two filters ratio
method. The temperatures evaluated with this method for two
non-flaring active region plasma were in the range from around 1.5
MK to about 5 MK.

\begin{acknowledgements}
{\it Hinode} is a Japanese mission developed and launched by
ISAS/JAXA, with NAOJ as domestic partner and NASA and STFC (UK) as
international partners. It is operated by these agencies in
co-operation with ESA and NSC (Norway).

This paper was supported by the Polish Ministry of Science and
Higher Education, grants N203 022 31/2991 and N203 015 32/1891.
\end{acknowledgements}

\end{document}